**Tittle**

# Prediction of Spherical Bubble-Chain-Induced Liquid Flow through Far-Wake Superposition Based on Bubble-Chain Hydrodynamics

**Author names**

Satoi SUZUKI [a], Hiroaki KUSUNO [b], Toshiyuki SANADA [a*]

**Affiliations**

[a] Department of Mechanical Engineering, Shizuoka University, Chuo-ku, Shizuoka 432-8561, Japan

[b]Department of Mechanical Engineering, Kansai University, Osaka 564-8680, Japan

***Correponding author**

Toshiyuki SANADA

sanada.toshiyuki@shizuoka.ac.jp

**Abstract**

Recent experimental observations revealed that clean spherical bubble chains generate a nearly uniform upward liquid flow. The physical origin of this liquid flow, however, remains unclear. Hydrodynamic interactions in aligned bubble chains were therefore investigated using high-accuracy embedded-boundary simulations that resolve the interfacial boundary layer while capturing long-range interactions among multiple clean spherical bubbles. The simulations were used to quantify the evolution of hydrodynamic interactions within bubble chains and to identify the conditions under which bubble-induced liquid flow can be represented by the superposition of isolated-bubble far wakes. Based on these findings, a reduced-order model was developed and applied to experimentally measured bubble trajectories. The model successfully reproduced the nearly uniform upward liquid flow observed in the experiments. To clarify the role of bubble dispersion, the predictions were compared with those for a uniformly dispersed bubble arrangement having the same overall dispersion width. While the uniformly dispersed arrangement produced a center-peaked velocity distribution, only the experimentally observed bubble trajectories reproduced the nearly uniform upward liquid flow. These results demonstrate that the liquid flow is governed not simply by the dispersion width but by the bubble trajectories that establish the spatial distribution of far wakes. The present study provides a physical framework linking hydrodynamic interactions among bubbles, bubble dispersion, and bubble-induced liquid flow, together with a reduced-order model for predicting the liquid flow generated by bubble chains.

**Highlights**

Interface-resolved DNS reveals hydrodynamic interactions in aligned bubble chains.

Far-wake superposition provides a reduced-order model for bubble-induced liquid flow.

A steady wake structure is established after only a few bubbles.

Two-stage bubble dispersion generates the nearly uniform upward liquid flow.

## 1. Introduction

Bubble-induced liquid flow plays a fundamental role in momentum transport, mixing, and mass transfer in gas–liquid flows [1]-[3]. Understanding the hydrodynamic interactions between bubbles and the surrounding liquid is therefore essential for predicting transport processes in bubble columns, chemical reactors, and many other engineering applications. Because practical bubbly flows consist of numerous interacting bubbles, identifying the physical mechanisms governing liquid transport directly from such complex systems remains challenging. A bubble chain provides the simplest canonical configuration in which the collective momentum transfer from multiple bubbles to the surrounding liquid can be isolated and systematically investigated, thereby offering fundamental physical insight into the liquid flow generated by bubble swarms. In addition, clean spherical bubbles have recently attracted renewed attention owing to the rapid development of electrochemical systems, such as water electrolyzers and fuel cells, where numerous small bubbles are continuously generated near electrode surfaces[4].

Liquid transport induced by a rising bubble results from both the Darwinian drift and the wake generated behind the bubble [5]-[8]. For deformable bubbles, enhanced surface vorticity generation leads to the formation of a standing eddy in the near wake, which significantly augments the liquid transport associated with the Darwinian drift. At higher Reynolds numbers, vortex shedding further intensifies the momentum exchange between the bubble and the surrounding liquid. Consequently, liquid transport induced by deformable bubbles has been extensively investigated in relation to their wake dynamics. In contrast, clean spherical bubbles do not generate a standing eddy at any Reynolds number, nor do they exhibit vortex shedding [9]-[11]. Consequently, the hydrodynamic mechanisms governing liquid transport around clean spherical bubbles remain much less understood than those around deformable bubbles.

To quantify the liquid transport induced by a single clean spherical bubble, researchers have developed theoretical and numerical descriptions focusing on the far wake. Batchelor [12] derived an analytical solution for the axisymmetric far wake of a moving body by assuming self-similarity in the momentum defect profile at large distances. Building on this theoretical foundation, Hallez and Legendre [13] performed detailed numerical simulations across various Reynolds numbers. They established a comprehensive formulation for the spatial decay and velocity distribution of the far wake behind an isolated clean spherical bubble, providing a fundamental physical basis for evaluating far-wake-induced liquid transport as follows:

$$u_{rot} = U \frac{C_{D\infty}\mathrm{Re}}{16} \frac{\alpha}{X} \exp\left[-\beta \frac{\mathrm{Re}}{8} \frac{Y^2}{X}\right], \tag{1}$$

where $C_{D\infty}$ is the drag coefficient of a single bubble, Re is Reynolds number defined by bubble terminal velocity $U$, $X$ and $Y$ is downstream and radial distance normalized by bubble radius. $\alpha$ and $\beta$ fit the theoretical solution to numerical result. The drag force acting on a clean spherical bubble characterized by the product of the drag coefficient and the Reynolds number ($C_D Re$) is strictly governed by the vorticity generated at the free-slip gas–liquid interface. Because this surface-generated vorticity is subsequently advected downstream without boundary layer separation to form the wake, the drag coefficient and the far-wake structure are inextricably linked.

Hydrodynamic interactions between clean spherical bubbles have been investigated extensively using bubble pairs as the simplest model system. Harper[14],[15] first demonstrated theoretically that the interaction results from a balance between wake-induced attraction and potential repulsion, predicting the existence of an equilibrium spacing between two bubbles. Subsequently, direct numerical simulations by Yuan and Prosperetti [16] quantified the drag reduction, rise velocity, and equilibrium spacing of two in-line bubbles, while Katz and Meneveau [17] investigated the same problem experimentally from the viewpoint of wake-induced interactions. Watanabe and Sanada [18] experimentally confirmed the existence of equilibrium distance at intermediate Reynolds numbers. Building on these studies, Ramírez-Muñoz et al. [19]-[23] demonstrated that the interaction forces acting on the trailing bubble are primarily governed by the far wake of the leading bubble, successfully reproducing the direct numerical simulations of Yuan and Prosperetti [16]. Hallez and Legendre [13] further generalized this concept into a unified interaction model capable of predicting both drag and lift forces for arbitrary bubble configurations. Consequently, the hydrodynamics of bubble pairs are now reasonably well understood in terms of far-wake-mediated interactions. More recently, advances in interface-resolved three-dimensional simulations have provided increasingly detailed insight into wake structures, bubble deformation, and multi-bubble interactions [24]-[27]. However, these simulations remain computationally demanding and have been largely limited to a small number of interacting bubbles. Consequently, whether the far-wake framework established for bubble pairs can be extended to predict the collective dynamics and the liquid flow induced by long bubble chains remains largely unexplored.

Despite the considerable progress in understanding pairwise interactions, the mechanism responsible for the liquid flow induced by bubble chains and their dispersion remains unclear. The dynamic behavior and stability of bubble chains have long been a subject of fundamental interest. Ruzicka [28] predicted bubble chain instability using a model that incorporates hydrodynamic interactions between bubbles. Sanada et al. [29] demonstrated that bubble chains diverged and scattered at high bubble generation frequencies and experimentally measured the surrounding liquid flow field induced by the chain. Atasi et al. [30] and Legendre and Zenit [31] mapped the stable and unstable regimes of bubble chains, identifying the wake-induced lift force as a key factor governing chain stability. Suzuki and Sanada [32] pointed out a strong two-way coupling between inter-bubble interactions, chain dispersion, and the mean upward liquid flow induced by the bubble chain itself.

They revealed that clean spherical bubble chains generate a nearly uniform upward liquid flow with a characteristic velocity reaching approximately 10% of the bubble rising velocity over a region extending several bubble diameters from the chain. The liquid flow is accompanied by a two-stage dispersion process, indicating a strong coupling between the collective bubble motion and the surrounding liquid [2]. Such a large-scale liquid flow cannot be explained by the near wake, whose influence decays rapidly with distance from an isolated bubble [13]. This naturally raises the question of whether the liquid flow can be explained solely by the superposition of the far wake generated by individual bubbles, or whether additional collective mechanisms are involved. Resolving this question is essential not only for understanding the hydrodynamics of bubble chains but also for establishing a predictive framework applicable to larger bubbly flows.

Because the drag coefficient and the far-wake structure of clean spherical bubbles are governed by the weak vorticity generated within the thin boundary layer along the free-slip gas–liquid interface, accurately resolving this boundary layer is essential [33],[34]. At the same time, the far wake extends over several tens of bubble radii, requiring computational domains spanning hundreds of bubble radii to investigate long bubble chains. This wide separation of spatial scales makes interface-resolved simulations particularly demanding. Previous studies have therefore relied primarily on boundary-fitted meshes [13],[35],[36] or highly refined interface-capturing methods [37],[38]. However, extending these approaches to long bubble chains remains computationally expensive because the boundary layer around every bubble must be sufficiently resolved over a large computational domain. On the other hand, Euler–Lagrange approaches provide an efficient framework for predicting large-scale bubbly flows using prescribed interaction models, but they are less suitable for investigating the physical mechanisms from which these interaction laws emerge [39][40]. These challenges motivate the development of an interface-resolved computational framework that combines adaptive mesh refinement with an efficient treatment of the gas–liquid interface, thereby preserving the essential bubble-scale physics while remaining applicable to long bubble chains.

To address these challenges, the present study consists of two complementary investigations. First, an interface-resolved simulation based on an embedded-boundary representation of free-slip spherical bubbles on an adaptive Cartesian mesh is developed to investigate hydrodynamic interactions within long bubble chains. This computational framework enables systematic analyses beyond the conventional two-bubble configuration and examines how the drag coefficient and the far-wake characteristics evolve with increasing bubble number[12],[13],[20],[41]. Second, based on the physical insights obtained from these simulations, a reduced-order model based on far-wake superposition is developed to predict the liquid flow induced by bubble chains. The proposed model provides a physics-based framework for examining whether the collective liquid flow generated by bubble chains can be explained by the superposition of individual far wakes, thereby bridging detailed interface-resolved simulations and predictive models for bubble-induced liquid flow.

## 2. Numerical methods

### 2.1 DNS around embedded bubbles in a chain

To quantify hydrodynamic interactions within bubble chains, direct numerical simulations (DNS) were performed for clean spherical bubbles aligned in the streamwise direction. Conventional boundary-fitted coordinate (BFC) methods become computationally demanding as the number of bubbles increases. Therefore, the open-source CFD solver Basilisk was employed [38],[42][43], in which spherical bubbles were represented using an embedded-boundary method [44],[45] with a free-slip boundary condition imposed on the bubble surface [46].

Figure 1 illustrates the computational domain. A cubic domain of $80R\times80R\times80R$, where $R$ is the bubble radius, was considered. Uniform flow entered from the left boundary and exited through the right boundary with an outflow boundary condition. On the remaining boundaries, Dirichlet boundary conditions corresponding to the free-stream velocity were imposed for the tangential velocity components. Up to $N=7$ spherical bubbles were embedded along the centerline of the computational domain. The streamwise direction was defined as the $x$-axis, while the transverse directions were denoted by the $y$- and $z$-axes.

The Reynolds number was varied from Re = 50 – 200. The center-to-center bubble spacing was varied over $S = 4 – 20$ to cover both near-field and far-field hydrodynamic interactions. In addition to perfectly aligned bubble chains, laterally staggered bubble arrangements were considered by varying the transverse displacement width from $W=2 – 6$ while maintaining the streamwise spacing. The computational cases are summarized in Table 1.

The flow was assumed to be incompressible and Newtonian. The governing equations are the incompressible continuity and Navier–Stokes equations,

$$\nabla \cdot \boldsymbol{u} = 0, \tag{2}$$

$$\frac{\partial \boldsymbol{u}}{\partial t} + \nabla \cdot (\boldsymbol{u} \otimes \boldsymbol{u}) = \frac{1}{\rho}[-\nabla p + \nabla \cdot (2\mu \boldsymbol{D})], \tag{3}$$

where $D = [\nabla u + (\nabla u)^T]/2$ is the deformation-rate tensor, $\boldsymbol{u}$ is the velocity, $p$ is the pressure, $\rho$ is the density, and $\mu$ is the dynamic viscosity [41],[47],[48]. The governing equations were discretized on a staggered Cartesian grid using second-order spatial discretization and a second-order fractional-step method for time integration.

To efficiently resolve both the interfacial boundary layer and long-range flow structures around multiple bubbles, adaptive mesh refinement (AMR) based on the embedded volume fraction and velocity field was employed. The maximum refinement level was set to level 12, corresponding to a minimum cell size of 0.0096$d$, where $d$ is the bubble diameter. Consequently, the bubble diameter was resolved by approximately 104 cells in the finest region near the bubble interface and in the wake.

The spherical bubble surfaces were represented using the embedded-boundary method. For a spherical interface, the free-slip boundary condition can be expressed as the following Neumann boundary condition,

$$\frac{\partial \boldsymbol{u}}{\partial n} = -\kappa \boldsymbol{u} \tag{4}$$

where $n$ is the outward normal direction and $\kappa$ is the mean curvature of the bubble surface. The grid convergence and the accuracy of the embedded free-slip boundary condition are assessed in Appendix A.

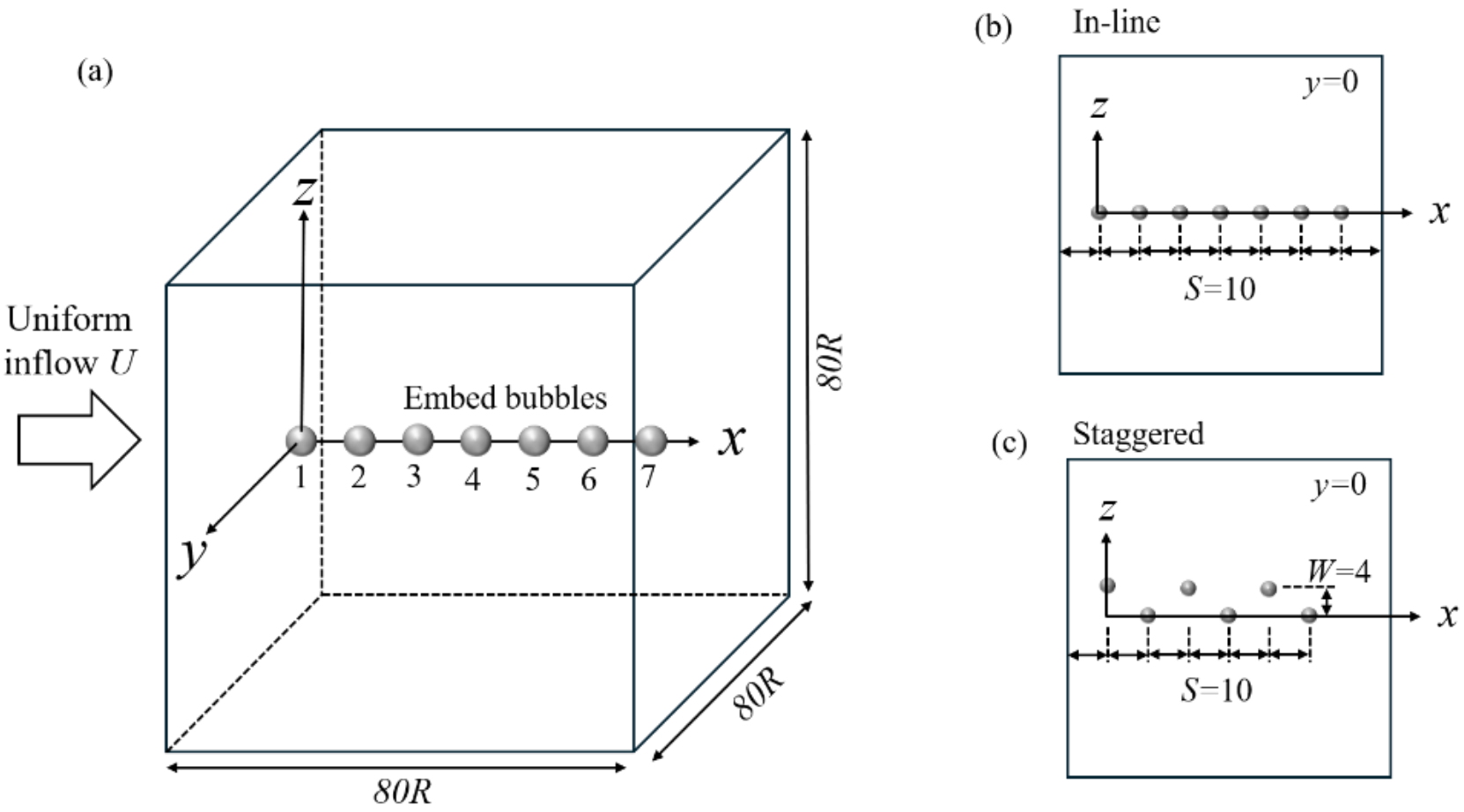

**Fig. 1 (a) The computational domain of embed bubbles in a chain and examples of bubble configuration of (b) in-line and (c) staggered.**

**Table 1. Numerical conditions.**

| Case | Re [-] | $N$ [-] | $S$ [-] | $W$ [-] | Chain condition |
|---|---|---|---|---|---|
| 1 | 50 | 7 | 4 | 0 | In-line |
| 2 | 50 | 7 | 10 | 0 | In-line |
| 3 | 50 | 5 | 15 | 0 | In-line |
| 4 | 50 | 3 | 20 | 0 | In-line |
| 5 | 75 | 7 | 10 | 0 | In-line |
| 6 | 100 | 7 | 4 | 0 | In-line |
| 7 | 100 | 7 | 10 | 0 | In-line |
| 8 | 150 | 7 | 10 | 0 | In-line |

| 9 | 200 | 7 | 4 | 0 | In-line |
|---|---|---|---|---|---|
| 10 | 200 | 7 | 10 | 0 | In-line |
| 11 | 50 | 6 | 10 | 2 | Staggered |
| 12 | 50 | 6 | 10 | 4 | Staggered |
| 13 | 50 | 6 | 10 | 6 | Staggered |

**2.2 Reduced-order model for bubble-chain-induced liquid flow**

A reduced-order model was developed to predict the liquid flow induced by dispersed bubble chains. The model is based on the assumption that, once the bubbles become sufficiently separated, interactions among the individual wakes become weak, and the flow induced by each bubble can be approximated by the far wake of an isolated bubble. The validity of this assumption is discussed based on the numerical results presented in Section 3. Under this assumption, the liquid flow generated by a bubble chain is represented by the linear superposition of the far wakes produced by the individual bubbles.

The velocity distribution in the far wake of an isolated clean spherical bubble has been described through asymptotic analyses and numerical simulations by Hallez and Legendre [13] and Ramírez-Muñoz et al. [49] The present model combines these far-wake descriptions with bubble trajectories reconstructed from experimental observations to predict the liquid flow generated by dispersed bubble chains.

**2.2.1 Bubble trajectory model**

The bubble trajectories used in the reduced-order model were reconstructed from the experimentally measured dispersion characteristics of clean spherical bubble chains reported by Suzuki and Sanada [32]. The experiments showed that the bubble chain forms an approximately ellipsoidal dispersion envelope whose lateral extent gradually increases with the distance from the bubble generator. The dimensions of the envelope are characterized by the semi-major axis $a(z)$ and the semi-minor axis $b(z)$, both of which vary with the vertical position $z$. Representative values of $a$ and $b$ are summarized in Table 2.

To examine the influence of the bubble distribution on the induced liquid flow, two trajectory models were considered. The first assumes that the bubbles are uniformly distributed within the experimentally determined dispersion envelope. The second reproduces the experimentally observed two-stage dispersion pattern, in which the bubbles initially separate into two branches and subsequently migrate along the outer boundary of the dispersion envelope. The two reconstructed trajectories are illustrated in Figs. 8 (a) and 9 (a), respectively.

**2.2.2 Far-wake superposition model**

Based on the reconstructed bubble trajectories described in the previous section, the liquid flow induced by a dispersed bubble chain was calculated by linearly superposing the far wakes generated by the individual bubbles. The computational procedure is illustrated schematically in Fig. 2. At each time step, the velocity induced by each bubble was evaluated at every computational point, and the total liquid velocity was obtained by summing the contributions from all upstream bubbles. The vertical liquid velocity is expressed as

$$U_z(x,y,z,t) = \sum_{i=1}^{N} U_{z,\mathrm{single}}(x - x_i(t), y - y_i(t), z_i(t) - z) \tag{5}$$

where $U_{z,\mathrm{single}}$ denotes the vertical velocity induced by an isolated bubble located at $(x_i(t), y_i(t), z_i(t))$, and $N$ is the number of upstream bubbles contributing to the velocity at the evaluation point.

In the present model, the vertical velocity induced by an isolated bubble is represented by combining a potential-flow contribution with an asymptotic far-wake contribution. The potential-flow component accounts for the inviscid disturbance near the bubble, whereas the far-wake component represents the long-range viscous disturbance associated with the momentum deficit. The latter is based on the self-similar far-wake solution proposed by Hallez and Legendre [13], which is consistent with the asymptotic wake description discussed by Ramírez-Muñoz et al. [49]. Defining the streamwise distance from the $i$-th bubble to the evaluation point as $s_i(t) = z_i(t) - z$ the vertical velocity induced by an isolated bubble is expressed as

$$U_{z,\mathrm{single}} = U\left(\frac{R}{s_i(t)}\right)^3 + \frac{UC_{D\infty}\mathrm{Re}}{16}\left(\frac{R}{s_i(t)}\right)\exp\left(-\frac{\mathrm{Re}}{8R}\frac{\left(x - x_i(t)\right)^2 + \left(\left(y - y_i(t)\right)\right)^2}{s_i(t)}\right) \tag{6}$$

where $U$ is the bubble rise velocity, $R$ is the bubble radius, Re is the Reynolds number, and $C_{D\infty}$ is the drag coefficient of an isolated bubble, evaluated using the correlation proposed by Mei et al. [50]. In this study, the bubble rising velocity was assumed to be constant at $U$ = 100 mm/s, with a Reynolds number Re = 50. That is, $s_i(t) = z_i(t) - z = Ut - z$. The first term of Eq.(6) represents the potential-flow disturbance, which decays rapidly with increasing distance from the bubble and therefore contributes mainly to the near-field flow. The second term represents the viscous far wake. Following the classical theory of self-similar wakes [12], the centerline wake velocity decreases approximately as $s_i^{-1}$, while the lateral wake width increases proportionally to $s_i^{1/2}$, owing to viscous diffusion of momentum. Because the objective of the present model is to predict the long-range liquid flow induced by dispersed bubble chains, the resulting flow is assumed to be governed primarily by the far-wake contribution.

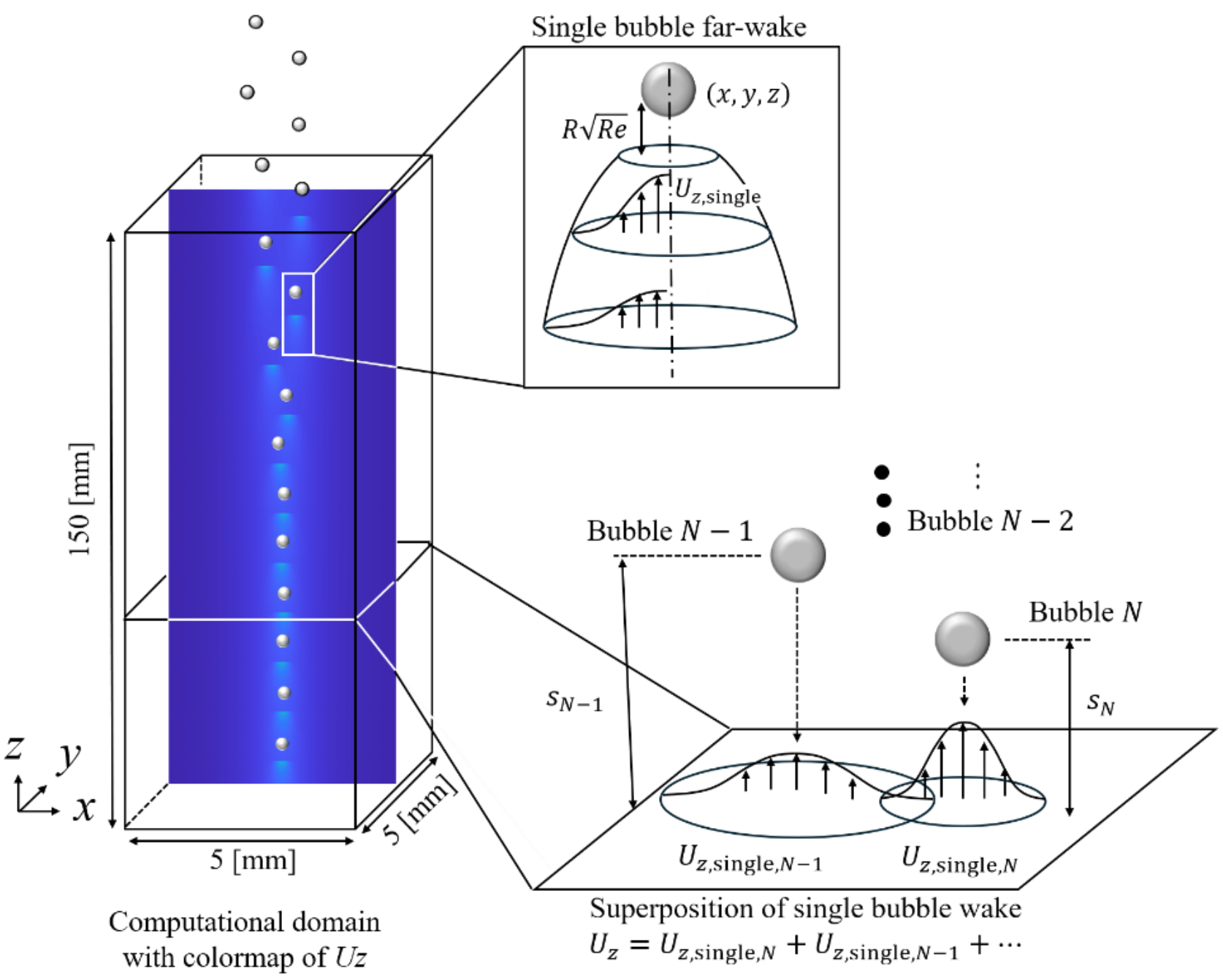

**Fig. 2 The schematic of upward flow model superimposing the wake of single bubble rising along the trajectory model.**

**2.2.3 Numerical implementation**

The liquid velocity field was evaluated within a computational domain of $10 \times 10 \times 150\ \mathrm{mm}^3$ centered on the bubble generation point. At each computational point, the velocity was obtained by summing the contributions from the upstream bubbles satisfying $s_i = z_i(t) - z > R\sqrt{Re}$. This criterion excludes the region close to each bubble, where the asymptotic far-wake expression is not applicable and becomes singular as $s_i$ approaches zero. The present calculation therefore focuses on the far-field liquid motion induced by the bubble chain rather than on the detailed near-wake flow.

The computational domain was discretized using a Cartesian grid. A spatial resolution of $\Delta x = \Delta y = 0.05$ mm was employed in the central region, $|x|, |y| \leq 3$ mm, whereas a spacing of $\Delta x = \Delta y = 0.1$ mm was used in the outer region. The grid spacing in the vertical direction was$\Delta z = 0.1$ m. The resulting computational grid contained approximately $3.9 \times 10^7$ evaluation points. The temporal evolution of the bubble chain was calculated from the generation of the first bubble. The simulation duration was 20 s for the low-frequency conditions, *f* = 4, 8Hz, and 10 s for the high-frequency conditions, *f* =12, 20Hz. The velocity fields obtained during these periods were used to evaluate the time-averaged liquid-flow distributions.

**Table 2. Dispersion range statistically obtained from experimental results.**

| $f$ [Hz] | $a$ (z=50 mm) | $b$ (z=50 mm) | $a$ (z=100 mm) | $b$ (z=100 mm) |
|---|---|---|---|---|
| 4 | 0.06 | 0.02 | 0.10 | 0.08 |
| 8 | 0.04 | 0.02 | 0.35 | 0.14 |
| 12 | 0.59 | 0.15 | 1.17 | 0.46 |
| 20 | 0.51 | 0.34 | 1.35 | 0.97 |

## 3. Hydrodynamic interactions in a bubble chain

### 3.1 Drag evolution in an aligned bubble chain

The hydrodynamic interaction in an aligned bubble chain was first examined by evaluating the drag coefficient acting on each bubble. Figure 3(a) presents the drag coefficient $C_{D,n}$ as a function of bubble number n for the reference condition, Re=50, S=10 and in-line configuration, together with the drag coefficient of an isolated bubble. The drag coefficient of the leading bubble agrees closely with the isolated-bubble value, confirming that the leading bubble is exposed to an essentially undisturbed upstream flow and that the present numerical method appropriately reproduces the drag acting on an isolated clean bubble.

In contrast, the drag coefficient decreases markedly for the second bubble because it is located within the wake generated by the leading bubble. As the bubble number increases further, the drag coefficient changes only slightly and becomes nearly independent of bubble number from the third bubble onward. Thus, the hydrodynamic interaction develops rapidly within the bubble chain and reaches an approximately stationary state after only a few bubbles.

The corresponding evolution of the wake structure is illustrated in Fig. 3(b), which shows the axial velocity-deficit field and the radial velocity-deficit profiles at x/R=3, 5, 6, and 8 downstream of each bubble. The wake behind the leading bubble differs from those behind the downstream bubbles, particularly in its magnitude and radial extent. However, the profiles behind the third and subsequent bubbles become nearly indistinguishable. This convergence of the downstream wake structure is consistent with the convergence of the drag coefficient shown in Fig. 3(a), indicating that the downstream bubbles are exposed to similar incoming flow conditions. For bubbles 1–6, the velocity deficit at $x/R$ = 8 exceeds those at $x/R$ = 4 and 6 because the trailing bubble induces an upward displacement effect.

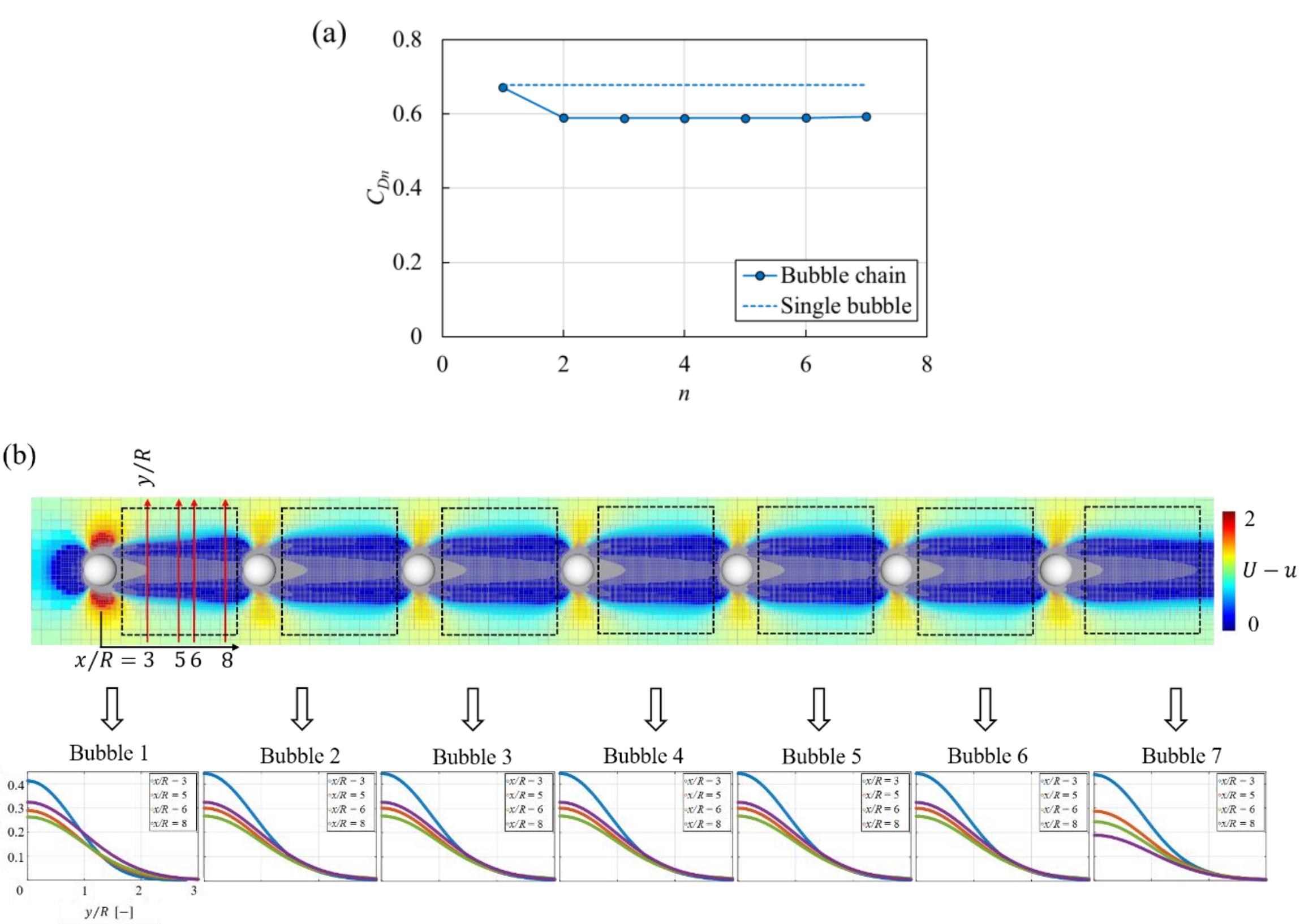

**Fig. 3. Hydrodynamic characteristics of an in-line bubble chain at Re=50 and S=10: (a) variation of the drag coefficient $C_{Dn}$ with bubble number n, and (b) radial profiles of the axial velocity deficit at x/R=3, 5, 6, and 8 downstream of each bubble. The drag coefficient and the downstream wake profiles become nearly independent of bubble number from the third bubble onward.**

The effect of the Reynolds number is shown in Fig. 4(a), where the drag coefficient is normalized by the isolated-bubble value. Although the magnitude of the drag reduction varies with Re, the same qualitative behavior is observed over the entire range investigated. The drag coefficient decreases substantially between the first and second bubbles and then remains nearly constant from the third bubble onward. The convergence of the normalized drag coefficient is therefore a robust feature of the aligned bubble chain rather than a behavior restricted to a particular Reynolds number.

Figures 4(b) and 4(c) show the effect of the bubble spacing at Re=50. Reducing the spacing strengthens the wake interaction and produces a larger reduction in the drag acting on the downstream bubbles. For the smallest spacing, S=4, the drag continues to decrease over the first several bubbles before gradually recovering near the downstream end of the finite chain. For the larger spacings, however, the drag coefficient rapidly reaches an approximately constant value. When plotted against the downstream distance from the leading bubble in Fig. 4(c), the results also show that the magnitude of the drag reduction is primarily determined by the bubble spacing, whereas the downstream variation becomes weak once the internal wake structure has developed.

These results demonstrate that the leading bubble behaves approximately as an isolated bubble, whereas the downstream bubbles experience a modified incoming flow created by the preceding wakes. Once this flow field becomes established, the drag and wake structure become nearly independent of bubble number. The physical origin of this common hydrodynamic environment is examined in the following section.

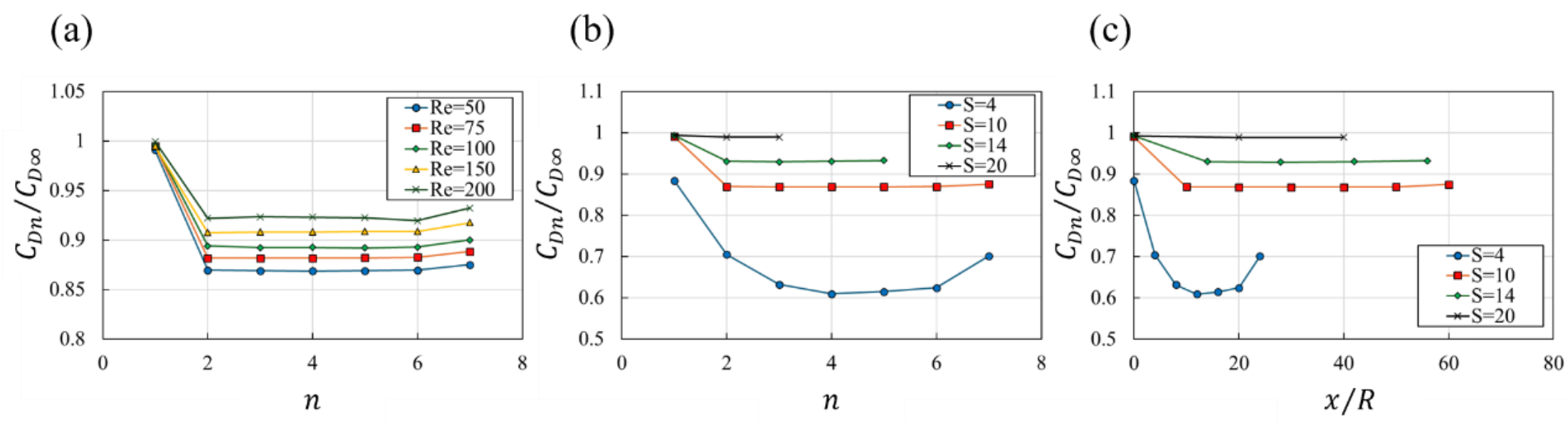

**Fig. 4. Effects of Reynolds number and bubble spacing on the drag coefficient in an in-line bubble chain: (a) normalized drag coefficient $C_{Dn}/C_{D\infty}$ as a function of bubble number n for different Reynolds numbers at S=10; (b) normalized drag coefficient as a function of bubble number for different bubble spacings at Re=50; and (c) normalized drag coefficient as a function of the downstream distance from the leading bubble for different bubble spacings at Re=50.**

### 3.2 Physical interpretation of drag convergence

The convergence of the drag coefficient suggests that the downstream bubbles are exposed to a periodically regenerated and nearly repeatable incoming flow. To clarify the physical origin of this behavior, the streamwise and radial characteristics of the wake are examined in Fig. 5. Figure 5(a) shows the streamwise variation of the centerline velocity deficit along the aligned bubble chain for different Reynolds numbers at S=10. A large velocity deficit is generated immediately downstream of each bubble and then decreases until the next bubble is encountered. The velocity deficit is subsequently regenerated behind that bubble, producing a repeated sequence of wake generation and decay. After the first few bubbles, this sequence becomes nearly periodic and only weakly dependent on bubble number. This periodic regeneration of the wake is consistent with the nearly constant drag acting on the downstream bubbles.

For comparison, the dashed line in Fig. 5(a) indicates the $X^{-1}$ decay characteristic of an isolated-bubble far wake [13],[49]. The centerline deficit within each interbubble region decreases in a manner broadly consistent with this algebraic decay, although it is interrupted and regenerated by the next bubble. Therefore, the downstream flow in an aligned chain may be regarded as a sequence of far-wake-like regions repeatedly generated by the individual bubbles.

The theoretical radial profiles of the isolated-bubble far wake at $x/R$=10 are shown in Fig. 5(b) [13]. The magnitude of the centerline velocity deficit varies only weakly with Re, whereas the radial width of the wake decreases as the Reynolds number increases. This behavior follows from the self-similar far-wake solution, in which the magnitude and spatial extent of the wake are governed by the product $C_D$Re [12],[49]. This product characterizes the momentum deficit carried by the wake and is related to the total vorticity generated at the bubble surface [35]. The relatively weak variation of $C_D$Re over the present Reynolds-number range therefore explains why the centerline velocity deficit is only weakly dependent on Re, even though the radial width of the wake varies appreciably.

The velocity at a single point on the centerline, however, does not directly represent the incoming flow responsible for the drag acting on a finite-sized bubble. The bubble interacts with the velocity distribution over its entire frontal projected area. The area-averaged axial velocity deficit is therefore evaluated in Fig. 5(c). As in the centerline results, the area-averaged deficit is regenerated downstream of every bubble and approaches an almost periodic distribution after the first few bubbles. The area average decreases with increasing Reynolds number because the far wake becomes narrower relative to the bubble diameter, even though the centerline deficit remains similar.

The corresponding effective Reynolds number $\mathrm{Re}_{\mathrm{eff}} = (U - \bar{u})d/\nu$, calculated from the area-averaged incoming velocity, is shown in Fig. 5(d). The effective Reynolds number is substantially lower than the bulk Reynolds number Re immediately behind each bubble because of the large local velocity deficit. It then increases downstream as the wake decays, before being reduced again by the next bubble. After the first few bubbles, the distribution of $\mathrm{Re}_{\mathrm{eff}}$ becomes nearly periodic. Therefore, each downstream bubble is exposed to approximately the same effective incoming Reynolds number, providing a direct explanation for the convergence of its drag coefficient.

The physical mechanism may consequently be summarized as follows. The leading bubble generates a wake that reduces the incoming velocity experienced by the second bubble. This reduction lowers the effective Reynolds number $\mathrm{Re}_{\mathrm{eff}}$ and, consequently, the drag coefficient $C_D$ of the second bubble. Once the chain is sufficiently developed, each downstream bubble generates a wake of nearly the same strength and encounters a similar area-averaged incoming velocity. The associated values of $C_D$Re, momentum deficit, and surface-vorticity generation are therefore expected to become nearly repeatable from one bubble to the next. The repeated generation of similar far-wake structures produces the common hydrodynamic environment responsible for the nearly constant downstream drag.

(a)

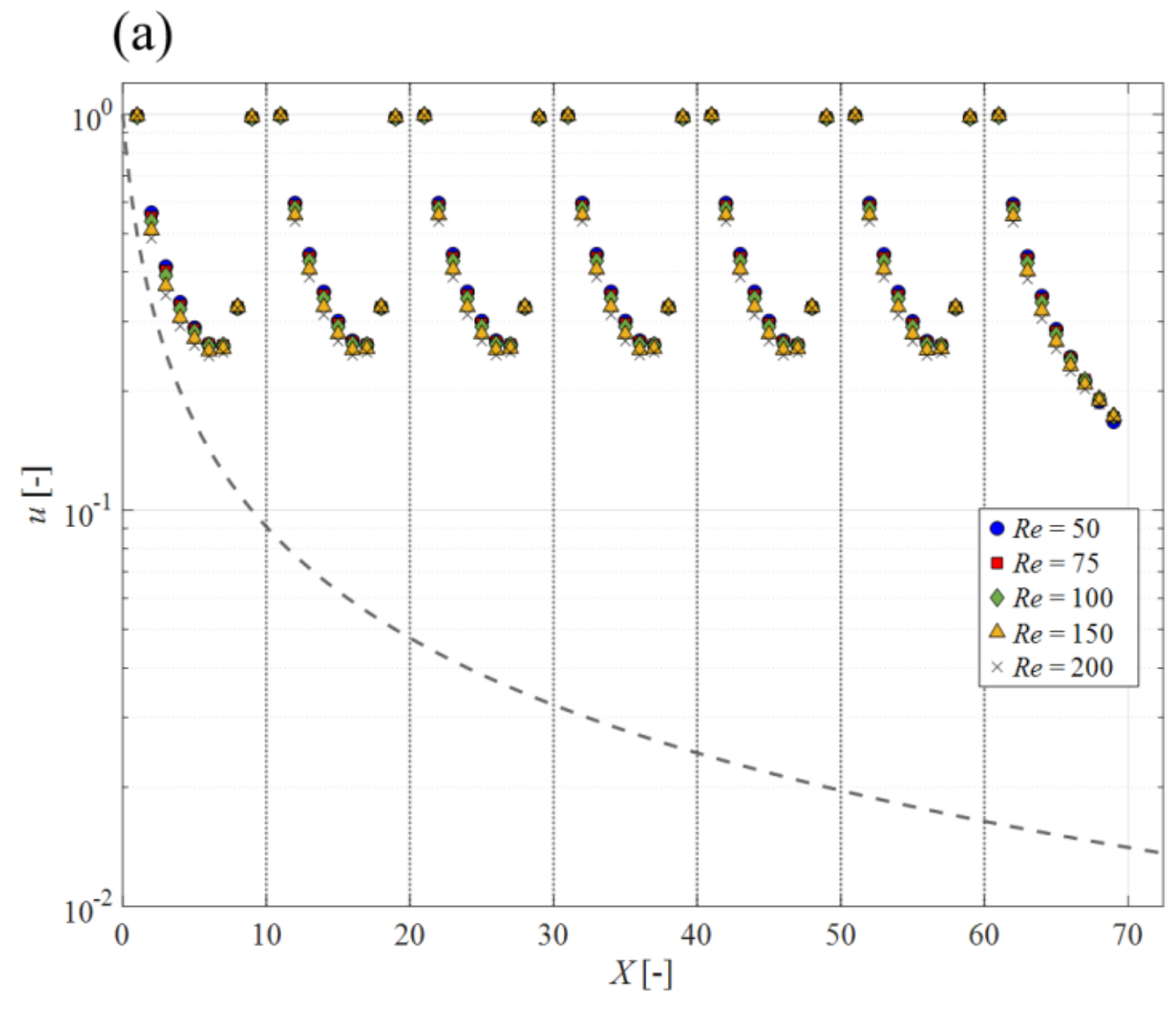

$10^0$
$10^{-1}$
$10^{-2}$
$u$ [-]
0
10
20
30
40
50
60
70
$X$ [-]
$Re = 50$
$Re = 75$
$Re = 100$
$Re = 150$
$Re = 200$

(b)

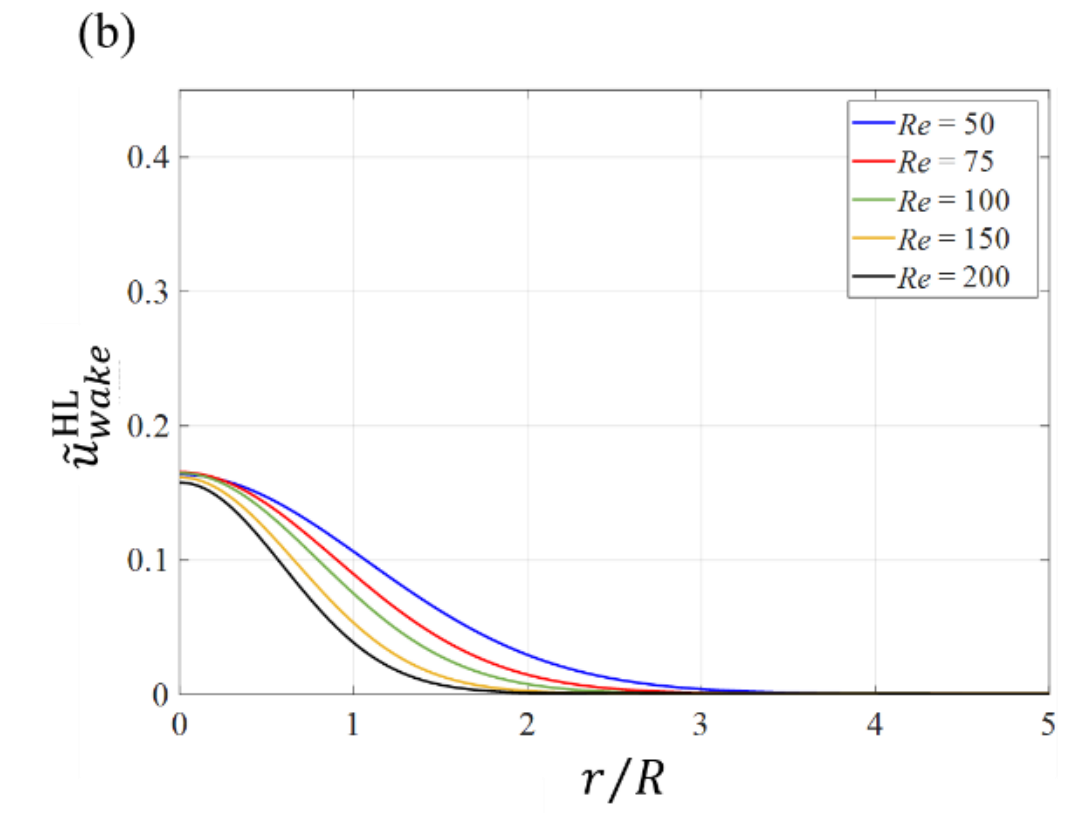

$\tilde{u}^{\mathrm{HL}}_{wake}$
0
0.1
0.2
0.3
0.4
0
1
2
3
4
5
$r/R$
$Re = 50$
$Re = 75$
$Re = 100$
$Re = 150$
$Re = 200$

(c)

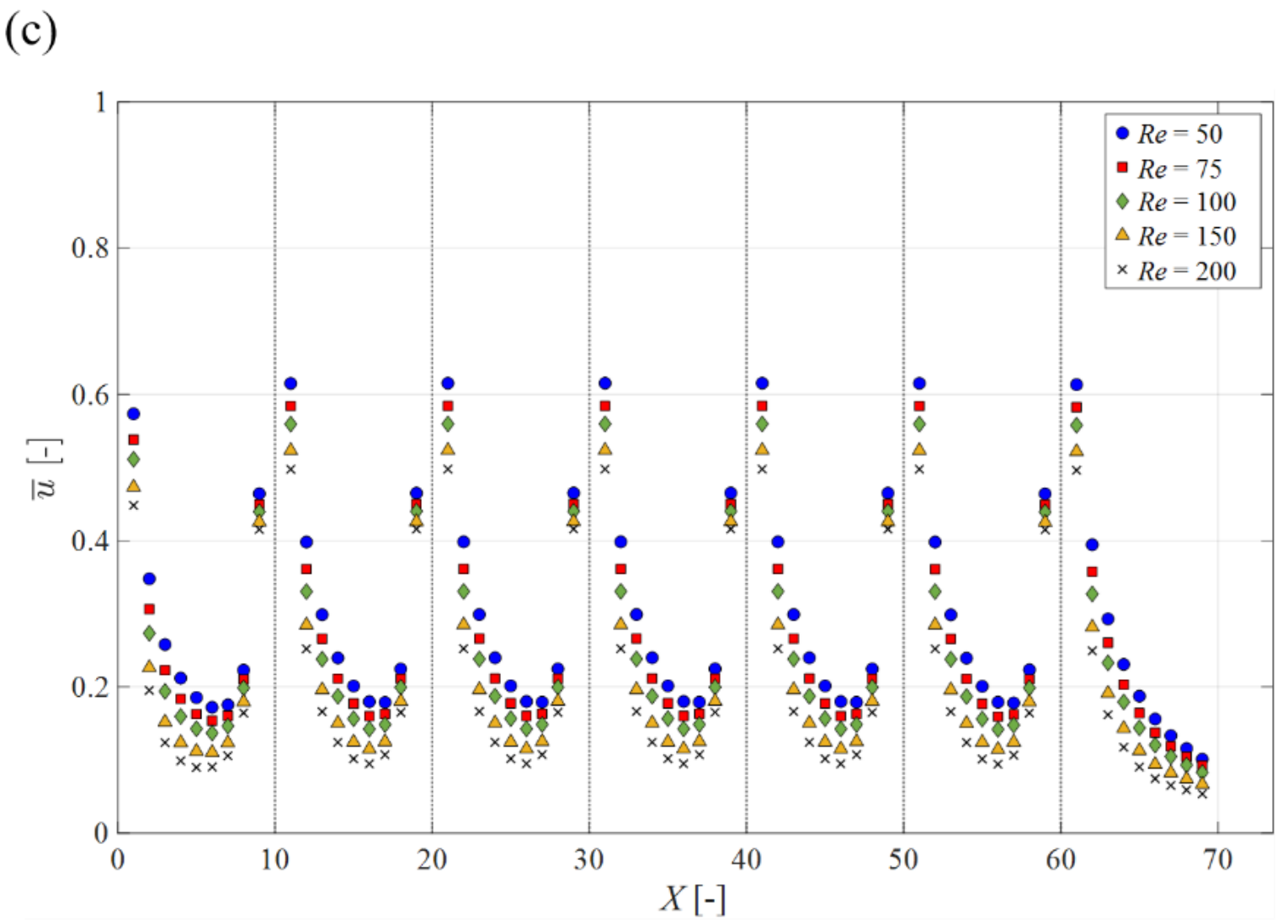

$\bar{u}$ [-]
0
0.2
0.4
0.6
0.8
1
0
10
20
30
40
50
60
70
$X$ [-]
$Re = 50$
$Re = 75$
$Re = 100$
$Re = 150$
$Re = 200$

(d)

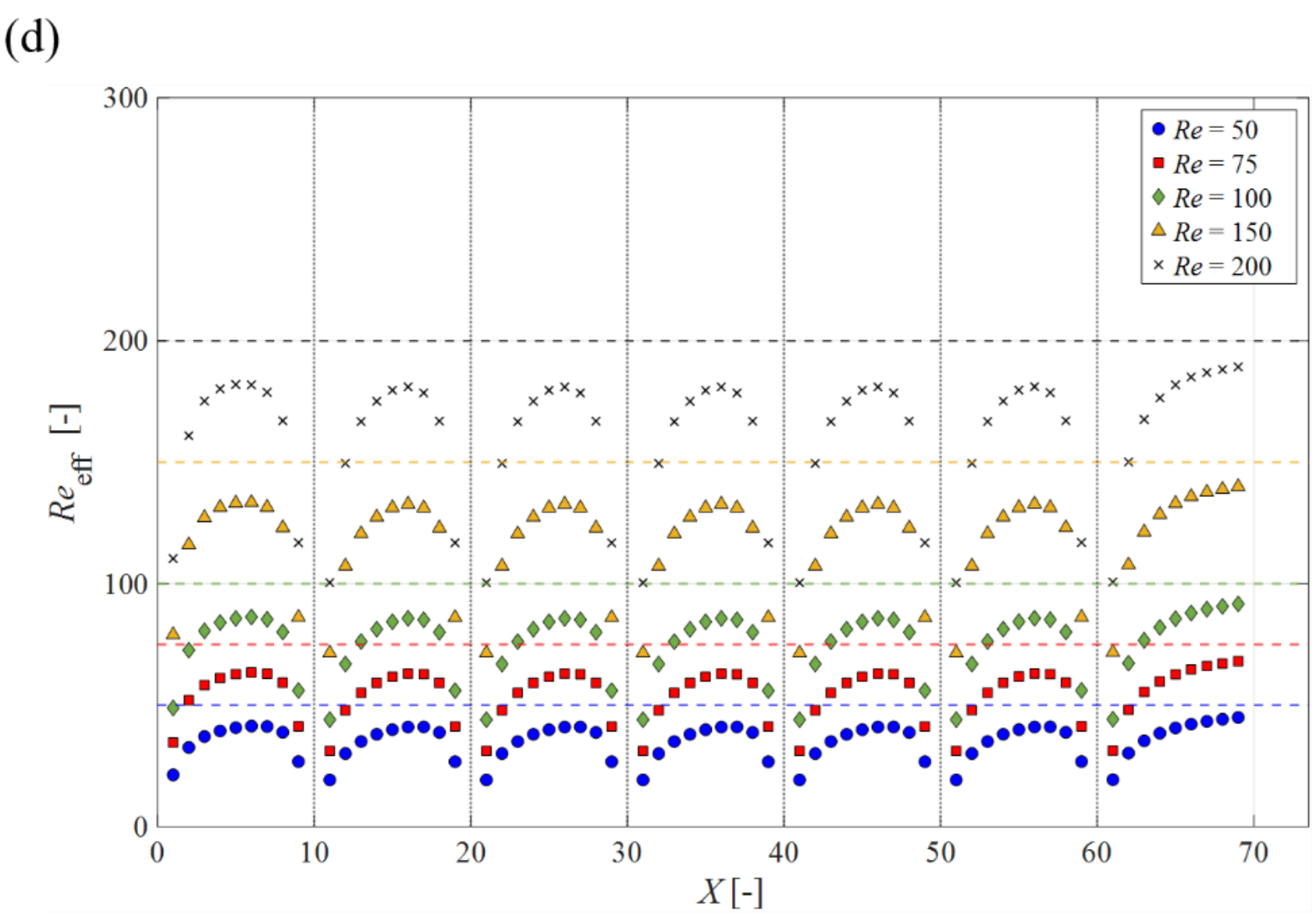

**Fig. 5. Evolution of the axial velocity deficit and the corresponding effective Reynolds number in an in-line bubble chain: (a) streamwise variation of the centerline velocity deficit for different Reynolds numbers at *S*=10, together with the $X^{-1}$ decay characteristic of the far wake; (b) radial profiles of the axial velocity deficit at *x/R*=10 behind an isolated bubble for different Reynolds numbers; (c) streamwise variation of the area-averaged axial velocity deficit over the frontal projected area of a bubble; and (d) effective Reynolds number $Re_{eff}$ evaluated from the area-averaged incoming velocity. The vertical bold lines indicate the bubble positions, and the horizontal lines in (d) indicate the corresponding bulk Reynolds numbers.**

### 3.3 Effect of lateral displacement on the far-wake structure

The preceding analyses considered a perfectly aligned bubble chain, in which every downstream bubble was located on the centerline of the wake generated by the preceding bubble. This arrangement represents an extreme condition of strong wake interaction. In an actual bubble chain, however, wake-induced lift displaces the bubbles laterally from the in-line configuration [30]. It is therefore necessary to determine how rapidly the wake generated by a downstream bubble recovers toward that of an isolated bubble as lateral dispersion progresses.

Additional simulations were performed for a laterally staggered six-bubble chain. The streamwise spacing was fixed at *S*=10, while the nondimensional lateral separation was varied as *W*=2, 4, and 6. These values correspond approximately to lateral displacements of one, two, and three bubble diameters, respectively.

Figure 8 shows the axial velocity-deficit field for *W*=4, together with the transverse velocity-deficit profiles *u* downstream of Bubbles 2, 4, and 6. The flow-field visualization reveals two important features. First, the wake generated behind each bubble remains locally identifiable. Second, the individual wakes overlap downstream and form a broad region of upward liquid motion throughout the bubble chain. Thus, the large-scale upward flow is not associated with a single continuously connected wake. Rather, it is formed through the spatial overlap of the wakes generated by the individual bubbles.

The transverse profiles in Fig. 6 contain a principal peak associated with the wake of the bubble from which the downstream distance is measured and a secondary peak associated with a bubble in the opposite lateral row. The magnitude of the secondary contribution generally decreases with downstream distance, whereas the principal wake retains a structure similar to that produced by an individual bubble. The profiles behind Bubbles 2, 4, and 6 are also similar, showing that a repeated wake pattern is established within the staggered chain.

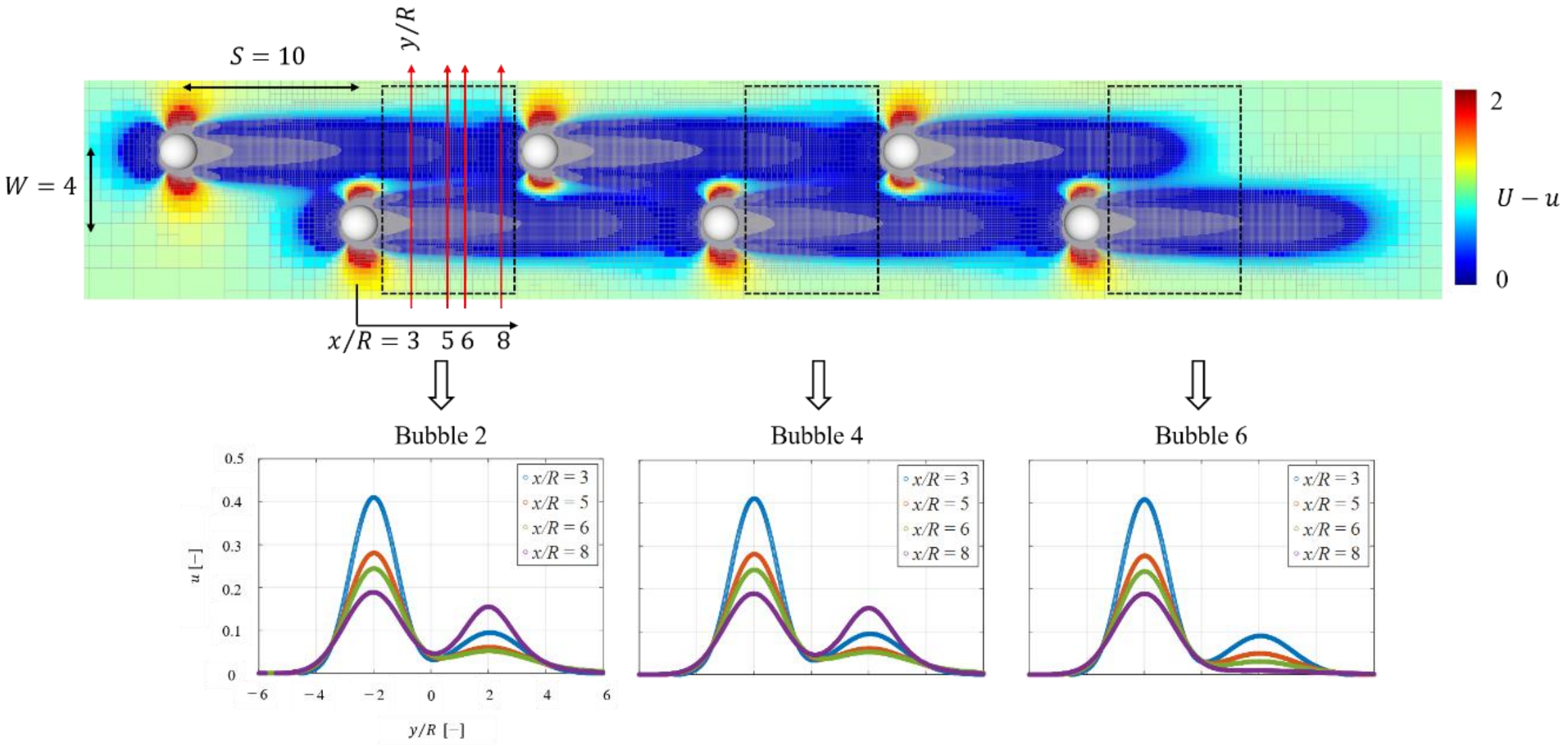

**Fig. 6. Wake structure in a laterally staggered six-bubble chain at Re=50, *S*=10, and *W*=4. The upper panel shows the axial velocity-deficit field, and the lower panels show the transverse profiles of the axial velocity deficit at *x/R*=3, 5, 6, and 8 downstream of Bubbles 2, 4, and 6.**

The effect of lateral separation is examined more directly in Fig. 7, where the profiles downstream of the sixth bubble are compared with the isolated-bubble far-wake solution. At *W*=2, corresponding to approximately one bubble diameter, the wakes of bubbles in the two lateral rows overlap strongly. The resulting velocity-deficit distribution is asymmetric and differs appreciably from the isolated-bubble solution. At *W*=4, the primary wake becomes more clearly separated from the wake generated by the opposite row, although a noticeable secondary contribution remains. At *W*=6, corresponding to approximately three bubble diameters, the primary wake profiles agree closely with the isolated-bubble far-wake solution over the downstream positions investigated. The wake generated by the displaced bubble can therefore be approximated by the isolated-bubble wake once the lateral separation reaches approximately three bubble diameters.

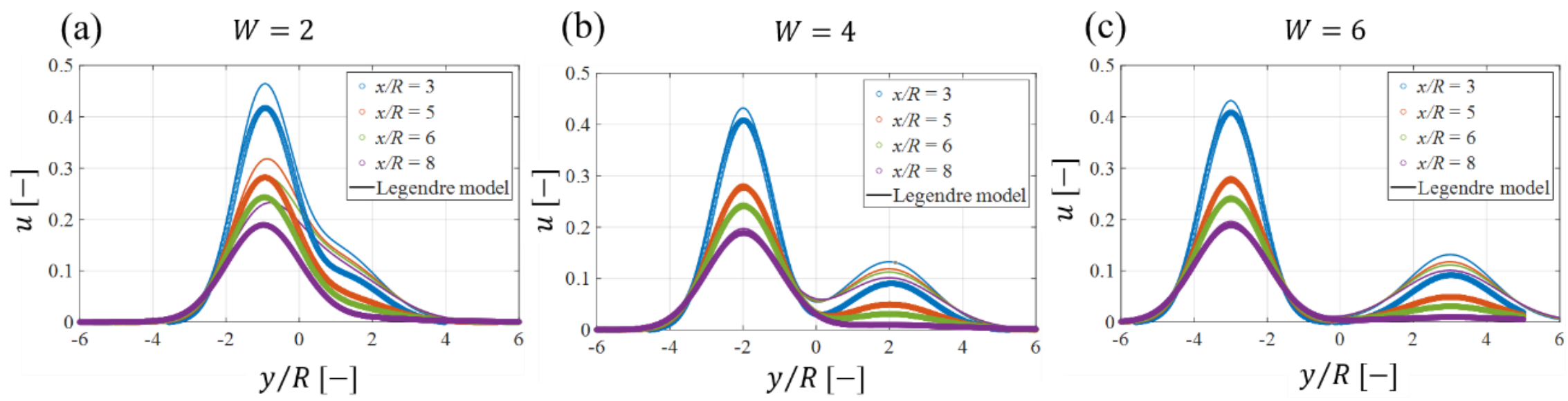

**Fig. 7. Transverse profiles of the axial velocity deficit downstream of the sixth bubble for nondimensional lateral separations of (a) *W*=2, (b) *W*=4, and (c) *W*=6 at Re=50 and S=10. Profiles are shown at *x/R*=3, 5, 6, and 8 downstream of the bubble and are compared with the isolated-bubble far-wake solution. With increasing lateral separation, the wake generated by the sixth bubble approaches the isolated-bubble wake, and the agreement is close at *W*=6.**

This rapid lateral recovery can be understood from the spatial scaling of the self-similar far wake [12],[49]. Its centerline velocity deficit decreases approximately as $X^{-1}$, while its characteristic radial width increases as $X^{1/2}$. The wake therefore persists over a long distance in the streamwise direction, but its influence remains confined to a comparatively limited lateral region at each downstream location. A bubble displaced by several diameters can consequently move outside the region of strong wake interaction, even though the weak upward velocity induced by the preceding wakes remains distributed over a large downstream distance.

This result is directly relevant to the two-stage dispersion mechanism identified in our previous study [32]. During the first stage, a relatively strong wake-induced lift drives a following bubble away from the wake centerline of the preceding bubble. Because this initial lateral migration is large, a displacement of several bubble diameters can be attained over a relatively short vertical distance. The bubbles therefore leave the region of strong in-line interaction before the slower second stage of dispersion becomes dominant.

After this initial lateral rearrangement, each bubble may still be transported by the combined velocity fields generated by the surrounding bubbles. Nevertheless, the far wake generated by each individual bubble becomes close to that of an isolated bubble. The large-scale liquid flow observed after dispersion can consequently be interpreted as the spatial superposition of individual isolated-bubble far wakes.

The present results thus provide the connection between the aligned-chain simulations and the modeling of a dispersed bubble chain. In the initial aligned configuration, strong pairwise wake interactions modify the drag and wake generated by the downstream bubbles. Wake-induced lift then rapidly separates the bubbles laterally, after which the individual wakes recover toward the isolated-bubble structure. On this basis, the following section develops a model in which the collective liquid flow induced by a dispersed bubble chain is represented by the linear superposition of isolated-bubble far-wake solutions.

## 4. Prediction of bubble-induced liquid flow by far-wake superposition

### 4.1 Validation of the far-wake superposition model

The far-wake superposition model described in Section 2 was applied to predict the liquid flow induced by a dispersed bubble chain. The statistical characteristics of bubble trajectories measured in the previous experimental study were used directly as the input to the model, while the liquid velocity field was reconstructed by linearly superposing the isolated-bubble far-wake solution. Since the objective of the present model is to reproduce the large-scale liquid motion, only the far-wake solution was considered, and the near-wake region was excluded from the superposition.

Figure 8(a) shows the reconstructed two-stage bubble trajectories together with the cross-sectional bubble distributions at several downstream locations. As reported previously, the dispersion develops in two distinct stages [32]. During the initial stage, the bubbles rapidly migrate away from the centerline owing to the wake-induced lift force generated by the preceding bubbles [13]. Subsequently, the dispersion proceeds more gradually while maintaining the annular distribution established during the first stage. The width of the dispersed region was extracted from the experimental results and used to reconstruct the two-stage trajectories.

The reconstructed liquid flow corresponding to these bubble trajectories is presented in Fig. 8(b). As the bubbles disperse laterally, the upward liquid flow gradually develops through the spatial superposition of the individual far wakes. Initially, only localized upward velocities are generated around the bubble trajectories. As the number of bubbles increases, however, these individual wakes overlap and form a continuous upward-flow region extending over the entire bubble chain.

The corresponding azimuthally averaged velocity profiles are shown in Fig. 8(c). At $z$ =50 mm, the upward velocity increases rapidly while maintaining a relatively broad profile. Further downstream, at $z$ =100 mm, the velocity distribution approaches a nearly flat profile over the central region. This characteristic feature agrees well with the experimental observations reported in our previous study, where a nearly uniform upward liquid flow was formed inside the dispersed bubble chain. Although the present model neglects the near-wake interactions and potential-flow effects, the large-scale liquid flow is reproduced with good agreement. These results indicate that the collective upward liquid flow is primarily governed by the spatial superposition of the far wakes generated by the dispersed bubbles.

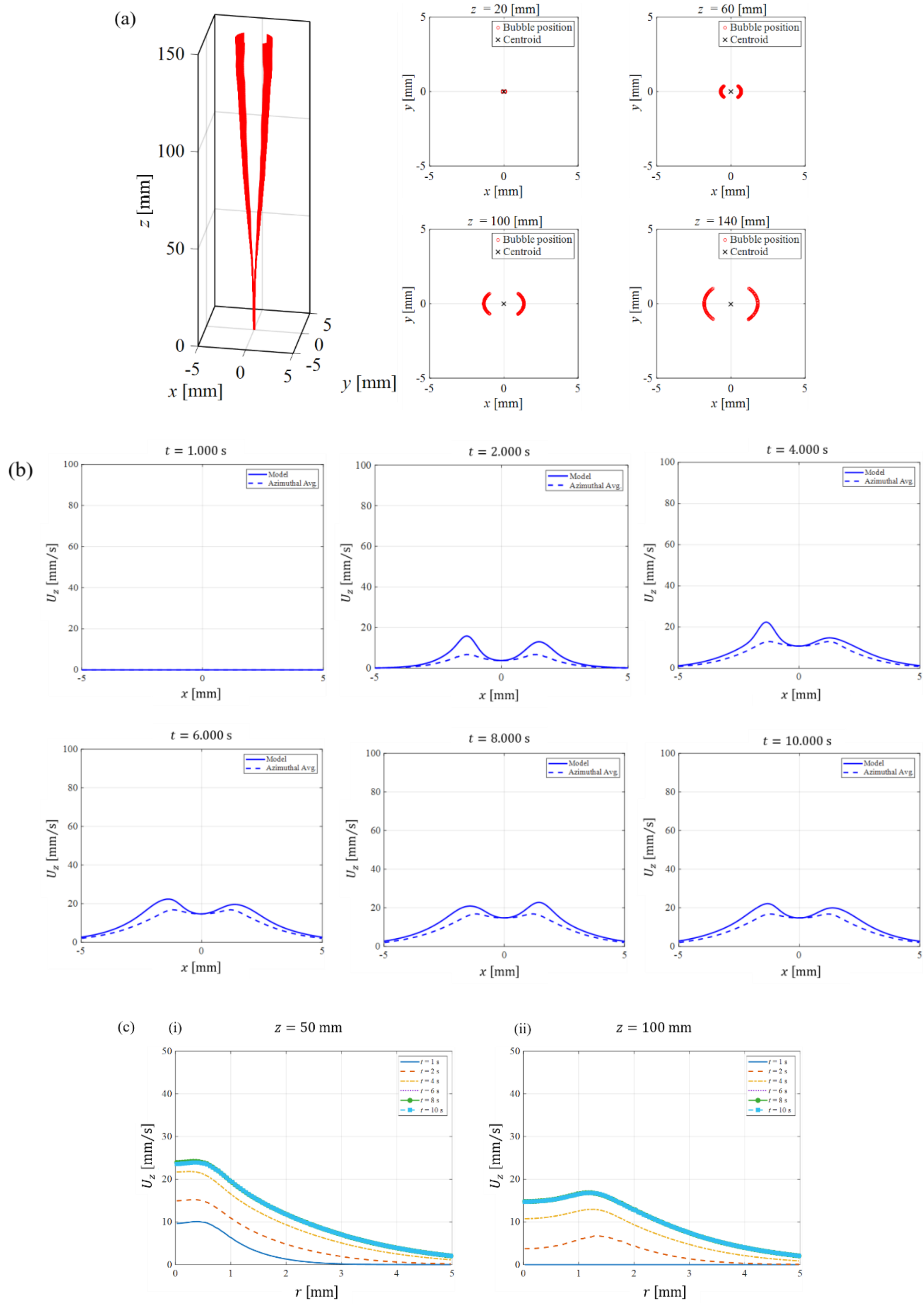

**Fig. 8 Trajectory of the two-stage dispersion model and development of the upward liquid flow reconstructed by the far-wake superposition. (a) Three-dimensional bubble trajectories and**

**cross-sectional bubble distributions at $z$ = 20, 60, 100, and 140 mm. (b) Evolution of the instantaneous axial velocity profile at $y$ = 0 and $z$ = 100 mm. Solid lines indicate the reconstructed velocity field, while dashed lines denote the azimuthally averaged velocity. (c) Evolution of the azimuthally averaged upward velocity profiles at (i) $z$ = 50 mm and (ii) $z$ = 100 mm. The bubble generation frequency is $f$ =20 Hz.**

### 4.2 Importance of the two-stage dispersion

To clarify the role of the two-stage dispersion in generating the upward liquid flow, a comparison was made with an artificial bubble arrangement in which the bubbles were uniformly distributed. The total number of bubbles and the overall dispersion radius were kept identical to those of the two-stage model so that only the spatial arrangement of the bubbles differed.

Figure 9(a) shows the uniformly dispersed bubble trajectories together with the corresponding cross-sectional distributions. Unlike the two-stage model, the bubbles remain distributed throughout the central region, producing an approximately uniform bubble density inside the dispersion radius.

The reconstructed liquid flow obtained from this bubble arrangement is presented in Fig. 9(b). In contrast to the two-stage model, the individual far wakes overlap predominantly around the centerline, producing a pronounced peak in the upward velocity. The azimuthally averaged velocity profiles shown in Fig. 9(c) exhibit a center-peaked distribution rather than the nearly uniform profile obtained for the two-stage dispersion model.

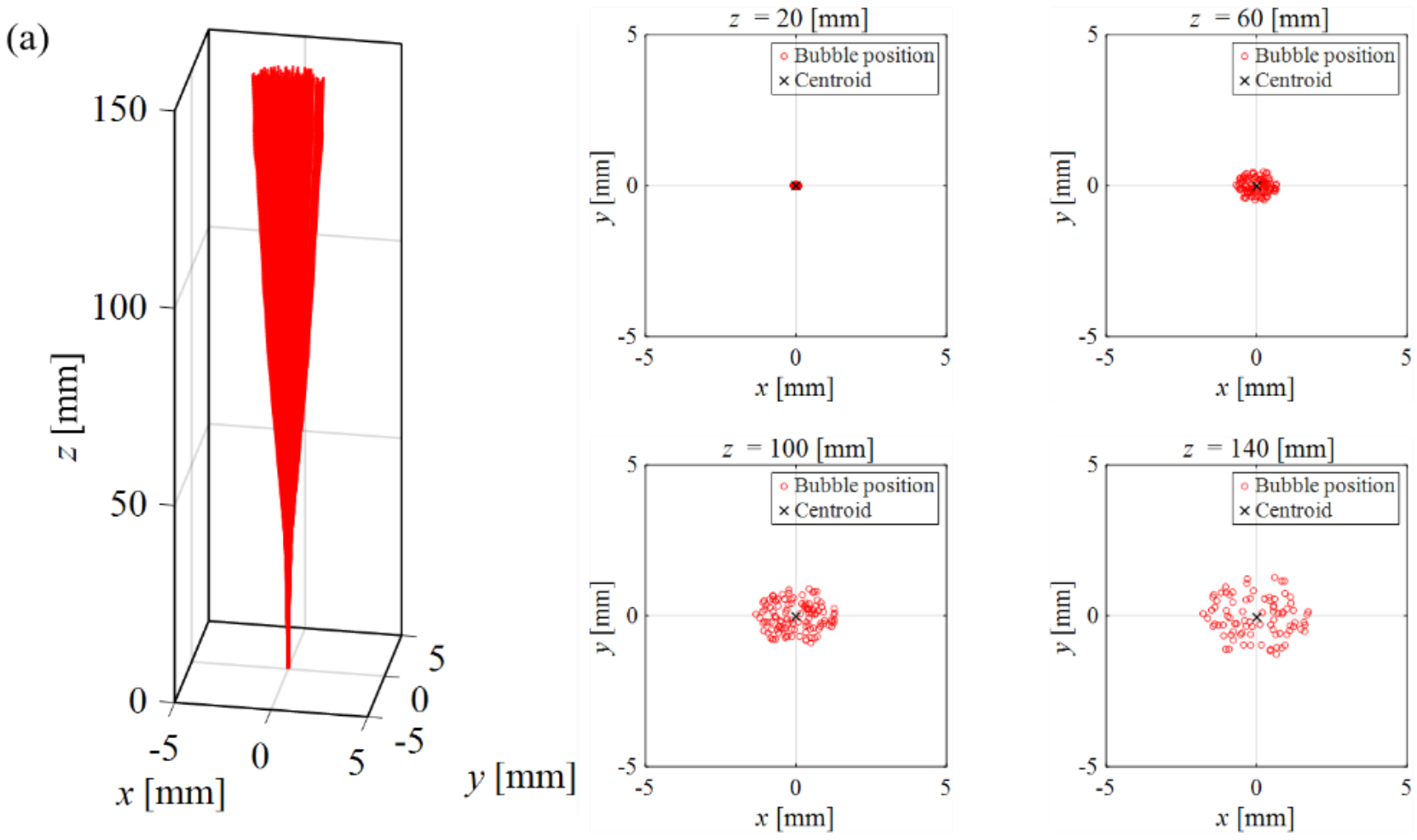

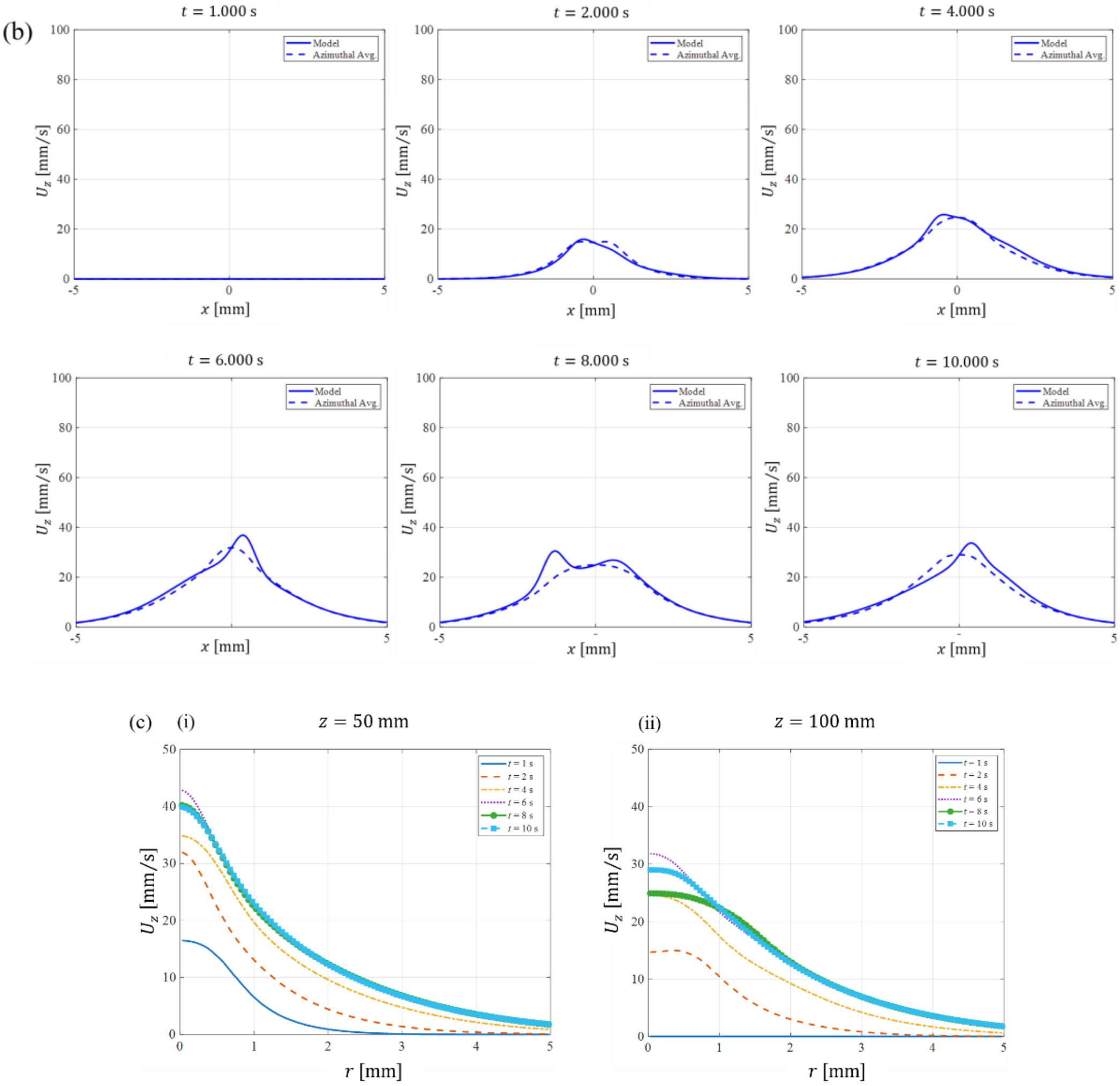

**Fig. 9 Trajectory of the uniformly dispersed bubble model and development of the upward liquid flow reconstructed by the far-wake superposition. (a) Three-dimensional bubble trajectories and cross-sectional bubble distributions at $z$ = 20, 60, 100, and 140 mm. (b) Evolution of the instantaneous axial velocity profile at $y$ = 0 and $z$ = 100 mm. Solid lines indicate the reconstructed velocity field, while dashed lines denote the azimuthally averaged velocity. (c) Evolution of the azimuthally averaged upward velocity profiles at (i) $z$ = 50 mm and (ii) $z$ = 100 mm. The bubble generation frequency is $f$ =20 Hz.**

The temporal evolution of the characteristic flow parameters is compared in Fig. 10. Although the uniformly dispersed model produces a larger maximum upward velocity, the characteristic width of the upward-flow region remains considerably smaller than that obtained from the two-stage model. In contrast, the two-stage dispersion generates a broader upward-flow region while maintaining a moderate maximum velocity. These results demonstrate that the overall dispersion width alone does not determine the liquid flow. Instead, the spatial arrangement of the bubbles plays a decisive role in shaping the upward velocity distribution.

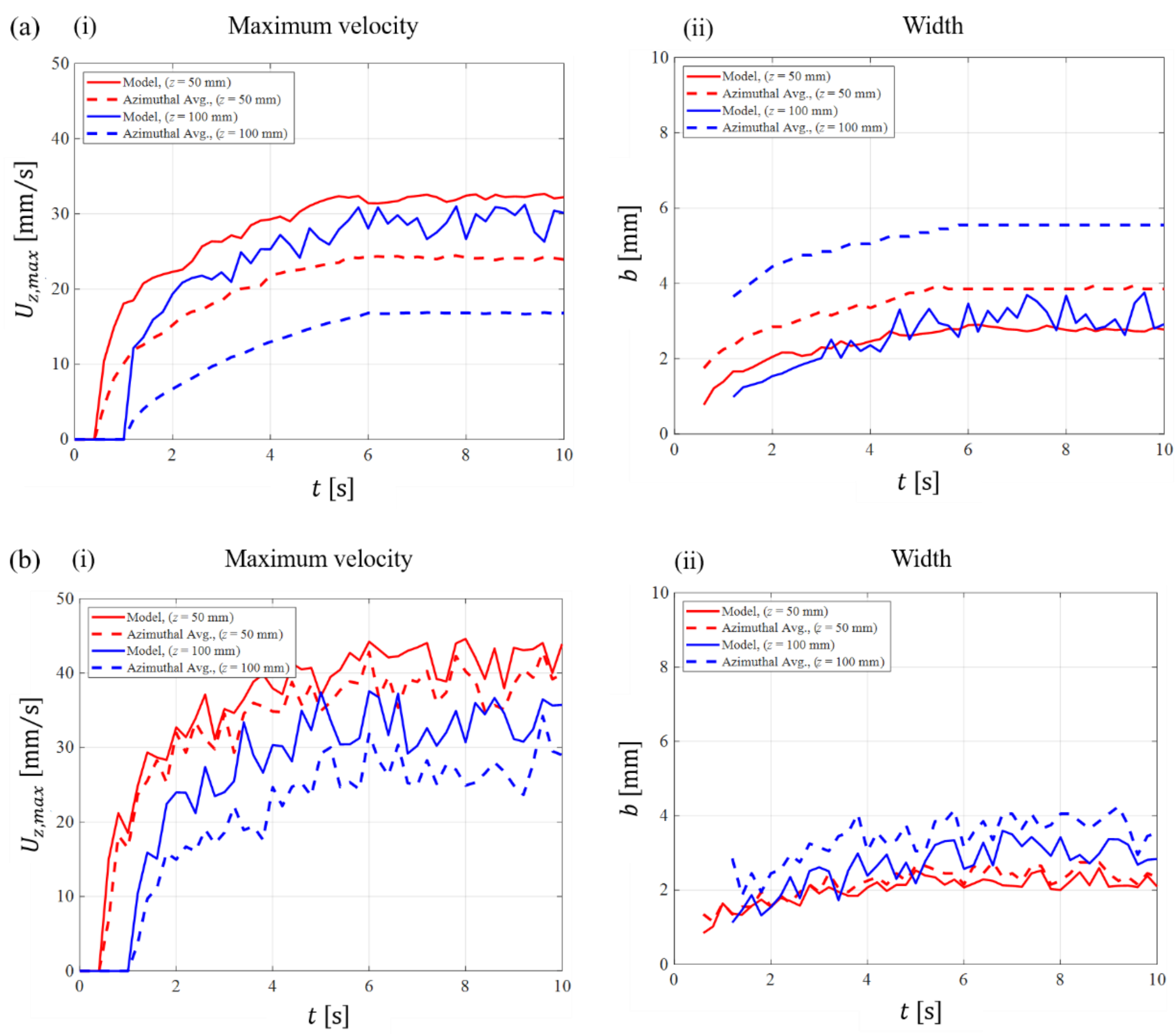

**Fig. 10 Temporal evolution of the characteristic upward-flow parameters reconstructed by the far-wake superposition for the two-stage dispersion model. Results at $z$ = 50 and 100 mm are compared. (a) two-stage dispersion model, (b) uniform trajectory model. (i) Maximum upward velocity. (ii) Characteristic width of the azimuthally averaged upward flow.**

The influence of the bubble generation frequency is examined in Fig. 11. As the generation frequency decreases, both the maximum upward velocity and the characteristic width decrease significantly. In particular, at $f$=4 Hz, the reconstructed velocity profile exhibits a distinct center peak, and the magnitude of the upward flow is only approximately 5% of the bubble rising velocity. This behavior agrees well with the experimental observations, where the bubble trajectories remained nearly linear and only a weak upward liquid flow was generated. As the generation frequency increases, the first stage of lateral migration becomes more pronounced, leading to a wider annular bubble distribution and consequently to the formation of the nearly uniform upward liquid flow [32],[30].

These comparisons demonstrate that the experimentally observed liquid flow cannot be explained solely by the overall dispersion of the bubble chain. Instead, the history of the dispersion process, particularly the rapid lateral migration during the first stage, determines the spatial arrangement of the wake sources and thereby controls the large-scale liquid flow.

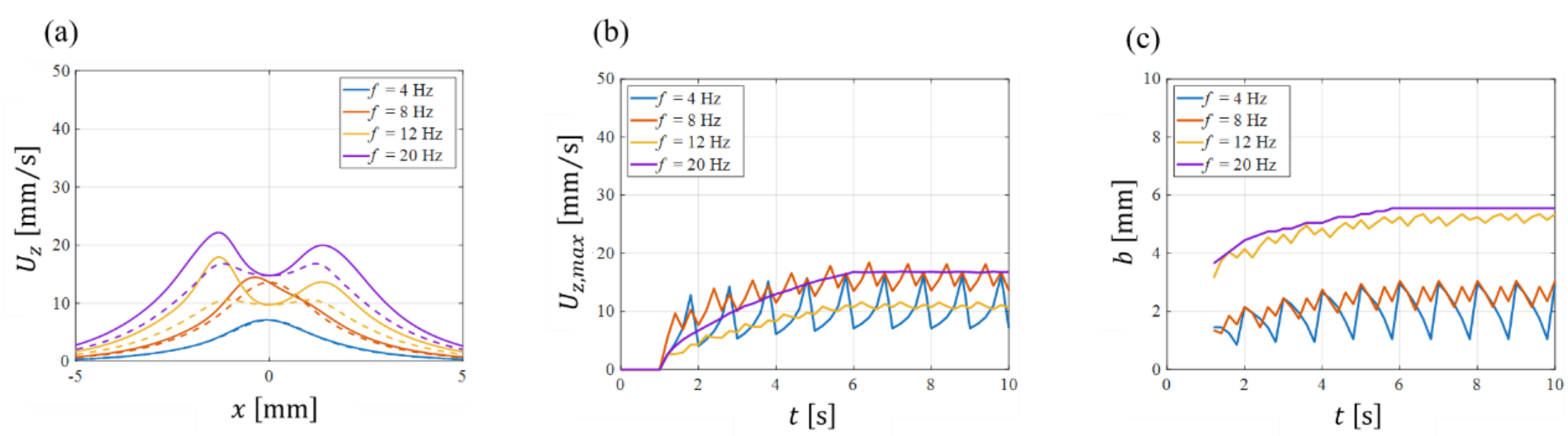

**Fig. 11 Effect of bubble generation frequency on the reconstructed upward liquid flow in the two-stage dispersion model. (a) Azimuthally averaged velocity profiles at z = 100 mm and t = 10 s. (b) Temporal evolution of the maximum upward velocity. (c) Temporal evolution of the characteristic width of the azimuthally averaged upward flow.**

### 4.3 Physical implications and future applications

The present results provide a physical interpretation of why the two-stage dispersion generates a nearly uniform upward liquid flow. During the first stage of the dispersion process, the wake-induced lift force rapidly transports the following bubbles away from the wake centerline of the preceding bubbles. Consequently, the wake sources become distributed around the central axis rather than remaining concentrated near the centerline. Since each dispersed bubble generates an isolated-bubble far wake, the superposition of these wakes naturally produces a broad and nearly uniform upward velocity distribution. In contrast, a uniformly dispersed bubble cloud concentrates the wake sources near the centerline, resulting in a center-peaked velocity profile.

Another important implication of the present study is that even small, clean, nearly spherical bubbles cannot necessarily be regarded as passive tracers. The disturbance generated by an individual bubble is relatively weak. However, the cumulative effect of many isolated far wakes produces a finite upward liquid flow whose magnitude reaches approximately 10% of the bubble rising velocity. The collective liquid motion therefore emerges from the accumulation of numerous weak disturbances rather than from strong interactions associated with individual bubbles.

Finally, the present framework suggests a possible reduced-order approach for Euler–Lagrange simulations of dispersed bubbly flows. The present model does not resolve the near-wake interactions, potential-flow effects, or other short-range bubble–bubble interactions. Instead, it focuses exclusively on the long-range influence of the isolated-bubble far wake. Despite this simplification, the model successfully reproduces the experimentally observed large-scale liquid flow and its dependence on the bubble arrangement and generation frequency. These results suggest that the far-wake superposition framework could provide an efficient reduced-order two-way coupling model, in which the carrier-flow field induced by dispersed bubbles is reconstructed without resolving the detailed flow around every individual bubble.

## 5. Conclusions

The present study investigated the hydrodynamic interactions within aligned bubble chains and the collective liquid flow induced by dispersed bubble chains by combining embedded-boundary numerical simulations with a far-wake superposition model. While previous studies have primarily focused on short-range interactions between neighboring bubbles or have treated small bubbles as passive particles in dispersed-flow simulations, the present study sought to establish a unified framework connecting local bubble interactions with the large-scale liquid flow generated by bubble chains. To this end, the hydrodynamic interactions in aligned bubble chains were first quantified using an embedded-boundary method with a slip boundary condition, and the induced liquid flow was subsequently reconstructed through the linear superposition of isolated-bubble far wakes. The results showed that the drag coefficient rapidly converges after the third bubble, that the wake generated by laterally dispersed bubbles approaches the isolated-bubble far wake, and that the experimentally observed upward liquid flow can be successfully reproduced by the far-wake superposition model.

These findings bridge the gap between the local hydrodynamics of individual bubbles and the collective liquid motion generated by bubble chains. In particular, the present study demonstrates that the large-scale liquid flow is governed not only by the overall dispersion of the bubbles but also by how the bubbles become dispersed. Comparison with an artificially generated uniformly dispersed bubble distribution revealed that the nearly uniform upward flow observed experimentally can only be reproduced when the experimentally observed two-stage dispersion process is considered. The first stage of wake-induced lateral migration distributes the wake sources around the centerline, enabling the superposition of isolated-bubble far wakes to produce the broad and nearly uniform upward liquid flow characteristic of bubble chains.

The present work therefore provides both a new physical interpretation of bubble-chain-induced liquid flow and a practical reduced-order modeling framework. Although the disturbance generated by each individual clean spherical bubble is relatively weak, the cumulative effect of many isolated far wakes produces a finite large-scale liquid flow whose magnitude reaches approximately 10% of the bubble rising velocity. This result suggests that even small, clean bubbles cannot necessarily be regarded as passive tracers in dispersed multiphase flows. More broadly, the proposed framework offers a physically based approach for representing the long-range feedback from dispersed bubbles to the carrier liquid without resolving the detailed flow around every individual bubble.

The present model is intentionally restricted to the far-wake region and neglects near-wake interactions, potential-flow effects, and other short-range hydrodynamic interactions between neighboring bubbles. Consequently, it is not intended to predict the detailed dynamics of closely spaced bubbles or bubble clustering. Future work should incorporate near-field interaction models into the present framework and couple the far-wake superposition model with Euler–Lagrange simulations. Such an extension would provide an efficient reduced-order two-way coupling approach capable of predicting large-scale bubble-induced liquid transport while preserving the essential physics governing dispersed bubbly flows. The present study provides a unified physical framework linking pairwise bubble interactions, two-stage bubble dispersion, and the generation of large-scale liquid flow in bubble chains.

**Acknowledgments**

This work was supported by JSPS KAKENHI Grant No. JP24K00801.

**Appendix A: Validation of the embeded boundary method**

To ensure the reliability and accuracy of the present numerical framework using the embedded boundary method in Basilisk [42], we conducted a validation study for a uniform flow past an isolated clean spherical bubble. The simulations were performed at Reynolds numbers of $Re = 50$ and $200$, and the results were evaluated in both the boundary layer and wake regions.

For the boundary layer validation, the computed velocity profiles were compared against results obtained from a Boundary-Fitted Coordinates (BFC) solver [51]. Figure A1 shows the radial distribution of the tangential velocity $u_t$ at $\theta = \pi/2$, where $\theta$ is the angle measured from the front stagnation point. For the velocity profiles obtained via the embedded boundary method, the values at the bubble surface ($r/R = 1$) were extrapolated from the fluid region using cubic spline interpolation. Figure A2 illustrates the tangential velocity distribution strictly along the bubble surface ($r/R = 1$). The results obtained using the embedded boundary method show excellent quantitative agreement with the BFC data. The simulations accurately capture the steep velocity gradients within the viscous boundary layer and confirm the precise implementation of the shear-free (full-slip) boundary condition on the spherical surface. The significant deviation from the theoretical potential-flow solution demonstrates that the viscous effects in the vicinity of the bubble are properly resolved.

Furthermore, we assessed the accuracy of the far-wake representation, which is a critical element for the far-wake superposition model discussed in Section 2.2. Figure A3 displays the transverse profiles of the wake velocity defect, $u_{wake}$, at various downstream locations ($x/R = 3,5,6,$ and $8$). The Basilisk simulation results are compared with the far-wake model proposed by Hallez and Legendre [13]. The computed wake profiles exhibit remarkable agreement with the analytical model at both $Re = 50$ and $Re = 200$, successfully predicting both the decay of the centerline velocity deficit and the lateral spread of the wake.

These validation results confirm that the present numerical setup appropriately resolves both the near-field boundary layer hydrodynamics and the far-field wake characteristics, thereby justifying the validity of the numerical simulations used in this study.

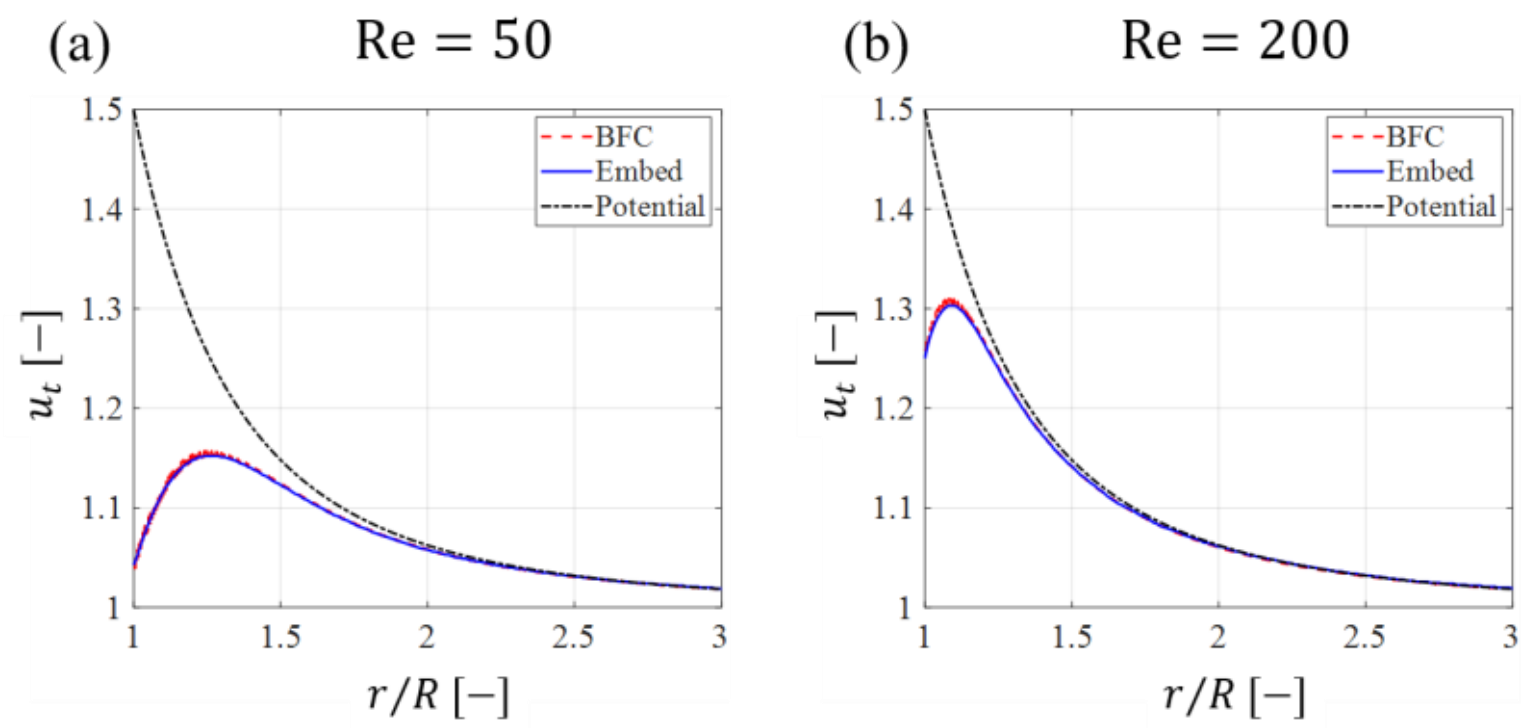

**Fig. A1 Tangential velocity profile around the boundary layer at (a) Re=50 and (b) Re=200. The angle from the front stagnant point $\theta = \pi/2$.**

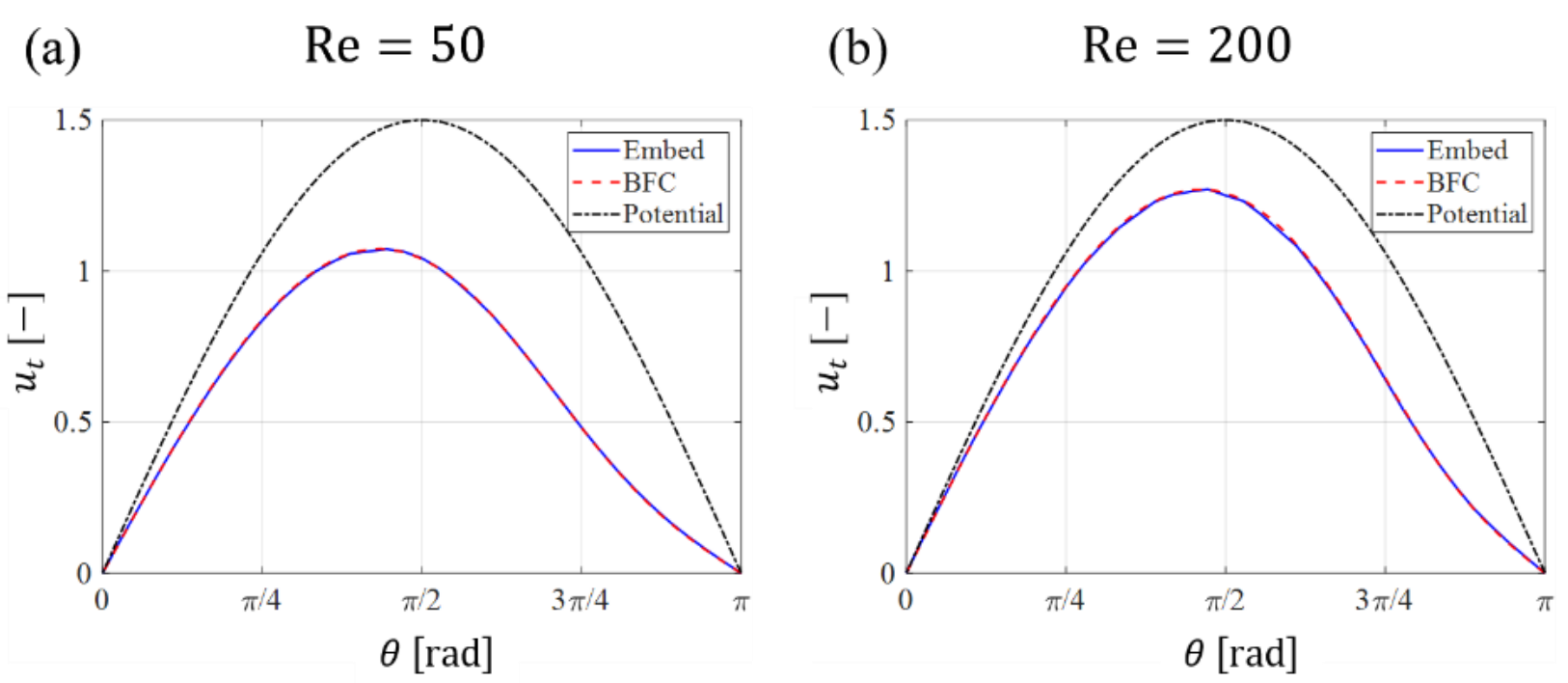

**Fig. A2 Tangential velocity profile as the function of the angle from the front stagnant point at (a) Re=50 and (b) Re=200.**

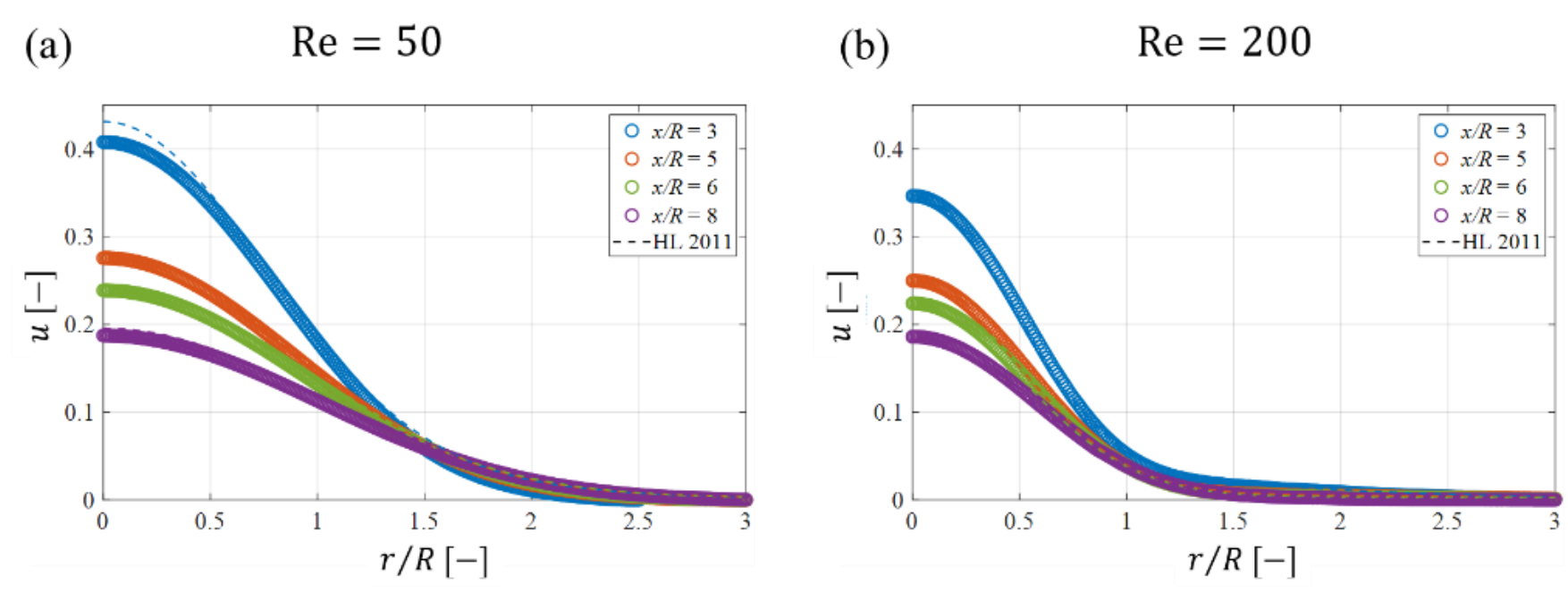

**Fig. A3 Velocity profile in the wake of a single bubble calculated by embed boundary code at**

**(a) Re=50 and (b) Re=200.**